\documentclass{article}
\usepackage{amsmath}

\usepackage{graphicx}

\begin{document}

\title{Resonating Delay Equation}


\author{Kenta Ohira\footnote{affiliated with a private company in Tokyo, Japan}}

\maketitle

\begin{abstract}%
We propose here a delay differential equation that exhibits a new type
of resonating oscillatory dynamics. The oscillatory transient dynamics
appear and disappear as the delay is increased between zero to asymptotically large delay. The optimal height of the power spectrum of the dynamical trajectory is observed with the suitably tuned delay. This resonant behavior contrasts itself against the general behaviors where an increase of delay parameter leads to the persistence of oscillations or more complex dynamics. 
\end{abstract}

\section{Introduction}

Studies of time delays in dynamics have found rather intricate
and complex behaviors. The time delays
most commonly occur due to finite conduction
and production times. They are intrinsic features of many
control and interacting systems and have been studied
in various fields including mathematics, biology, physics, engineering, and economics.\cite{heiden1979,bellman1963,cabrera_1,hayes1950,insperger,kcuhler,longtinmilton1989a,mackeyglass1977,miltonetal2009b,ohirayamane2000,smith2010,stepan1989,stepaninsperger,szydlowski2010}). 
``Delay Differential Equations'' are
the main mathematical approaches and modeling tools for
such systems. 

Typically,  delays induce instability of stable fixed points leading to
oscillatory and more complex dynamics. Also, the complexity of dynamics generally increases with longer delays. For example, the Mackey--Glass equation\cite{mackeyglass1977}, which was introduced to model the reproduction of the blood cells, 
shows the sequence of the monotonic convergence, transient oscillations, persistent oscillations, and chaotic dynamics as the delay parameter in the
feedback function becomes longer. Understandings of the path to the complex behaviors of many systems with delays, including this model, have been gradually gained (e.g.\cite{taylor}). There are, however,  more to be investigated and explored, particularly with respect to the nature of time-dependent dynamical trajectories.

Against this background, we propose here a delay differential equation that exhibits a new type of dynamical behavior. Namely, the oscillatory transient dynamics appear and disappear as the delay is increased between zero to asymptotically large delay. We analyze the equation both mathematically and numerically to show that there is a resonance: the optimal height of the power spectrum of the dynamical trajectory is observed with the suitably tuned value of the delay parameter. This resonant behavior contrasts itself against the general behaviors induced by delay where an increase of delay leads to the persistence or increase of complex dynamics.

We emphasize that we are not investigating stability switching phenomena (e.g.\cite{yan}) with the delay as the bifurcation parameter. Indeed, in our analysis of the proposed model in this work, the asymptotic stability of the fixed point never changes with increasing delay.  Changes are observed with the shapes of dynamical trajectories approaching the stable fixed point.

We end this paper with the discussion that there is an indication that similar transient resonant behaviors exist for more general types of equations.

\section{Main equation and its properties}

The equation we propose is the following:
\begin{equation}
{dX(t)\over dt} + a t X(t) = b X(t-\tau)
\label{dr}
\end{equation}
where $a \geq 0$, $b \geq 0$, $\tau \geq 0$ are real parameters.
When the parameter $\tau$ is interpreted as a delay, we can consider this equation as a delay differential equation describing the dynamics of the variable $X(t)$. 

We investigate this equation both analytically and numerically to show that oscillatory transient dynamics appear and disappear as the value of delay increases.

\subsection{Analysis of different cases}

Let us first consider the case that $b=0$. With the initial condition $X(t=0) = X_0$, the solution to the equation is given as
\begin{equation}
X(t) = X_0 e^{- {1\over 2}a t^2}
\label{sho}
\end{equation}
Thus, it is a trajectory with a gaussian shape. 

Next, the case that $a=0$ is a special case of much--studied Hayes's equation, 
first--order delay differential equation with constant coefficients\cite{hayes1950}. It is known that the origin $X=0$ is asymptotically stable only in the range of
\begin{equation}
- {\pi/{2 \tau} }< b <0.
\label{hayes}
\end{equation}

Now comes the general case with $a>0, b>0$. When the delay $\tau = 0$, the solution $X(t=0) = X_0$ is easily obtained as
\begin{equation}
X(t) = X_0 e^{- {1\over 2}a t^2 + b t}
\label{tau0}
\end{equation}
This is again a Gaussian with its peak at $b/a$.

To gain an insight when $\tau \neq 0$, we take the Fourier transform of the equation (\ref{dr}).
\begin{equation}
i\omega \hat{X}(\omega) + i a {d\hat{X}(\omega)\over d\omega} = - b \hat{X}(\omega) e^{i \omega \tau}
\label{ft}
\end{equation}
where
\begin{equation}
\hat{X}(\omega) = \int_{-\infty}^{\infty} e^{i \omega t} X(t) dt
\label{ft2}
\end{equation}
The solution is again readily obtained with ${\cal{C}}$ as the integration constant,
\begin{equation}
\hat{X}(\omega) = {\cal{C}} Exp[ {- {1\over 2 a} \omega^2 + {b\over \tau a} e^{i \omega \tau}}] 
\label{ftsol}
\end{equation}
From this expression, we can infer when $\tau$ becomes very large; the second term in the exponential approaches zero leading to the gaussian form. As we transform back, it is again the gaussian trajectory. Thus, with $\tau \rightarrow \infty$,
\begin{equation}
X(t) \rightarrow {\cal{D}}e^{- {1\over 2}a t^2}
\label{tau0}
\end{equation}
where ${\cal{D}}$ is a constant.

So, we have the gaussian trajectory with zero and asymptotically large delay. What happens in between? It turns out that oscillatory behaviors arise and disappear as we increase the value of the delay $\tau$. We will show these resonating phenomena together with numerical simulations
of the equation.

\subsection{Analysis of the Oscillatory Behavior}

We have performed the numerical simulation of equation (\ref{dr}). Some representative examples
are shown in Fig.1 with parameter choice so that the asymptotic stability is kept with any value of the delay. With zero delay, the shape of the dynamics is the gaussian as derived in the previous subsection. With the increasing delay, the oscillatory behaviors arise on top of the 
gaussian trajectory. Further increase changes the oscillatory shape into trains of pulses with decreasing height at the delay interval. In the limit of the long delay, the oscillation disappears. Note again that the asymptotic stability of $X=0$ does not change by the increasing delay in this parameter set. This is the one notable effect of having replaced the constant coefficient by the linearly dependent one on $t$ in the second term of the equation (\ref{dr}); it is in contrast to the case of much--studied constant coefficient case, where the onset of the oscillation leads to the destabilization by the increasing delay. It is also different from the delay induced transient oscillation (DITO)\cite{milton_mmnp,pakdamanetal1998a}. The phenomena arise in coupled delay differential equations exhibiting the prolonged duration of oscillatory behaviors with increasing delay. 

\begin{figure}
\includegraphics[height=16cm]{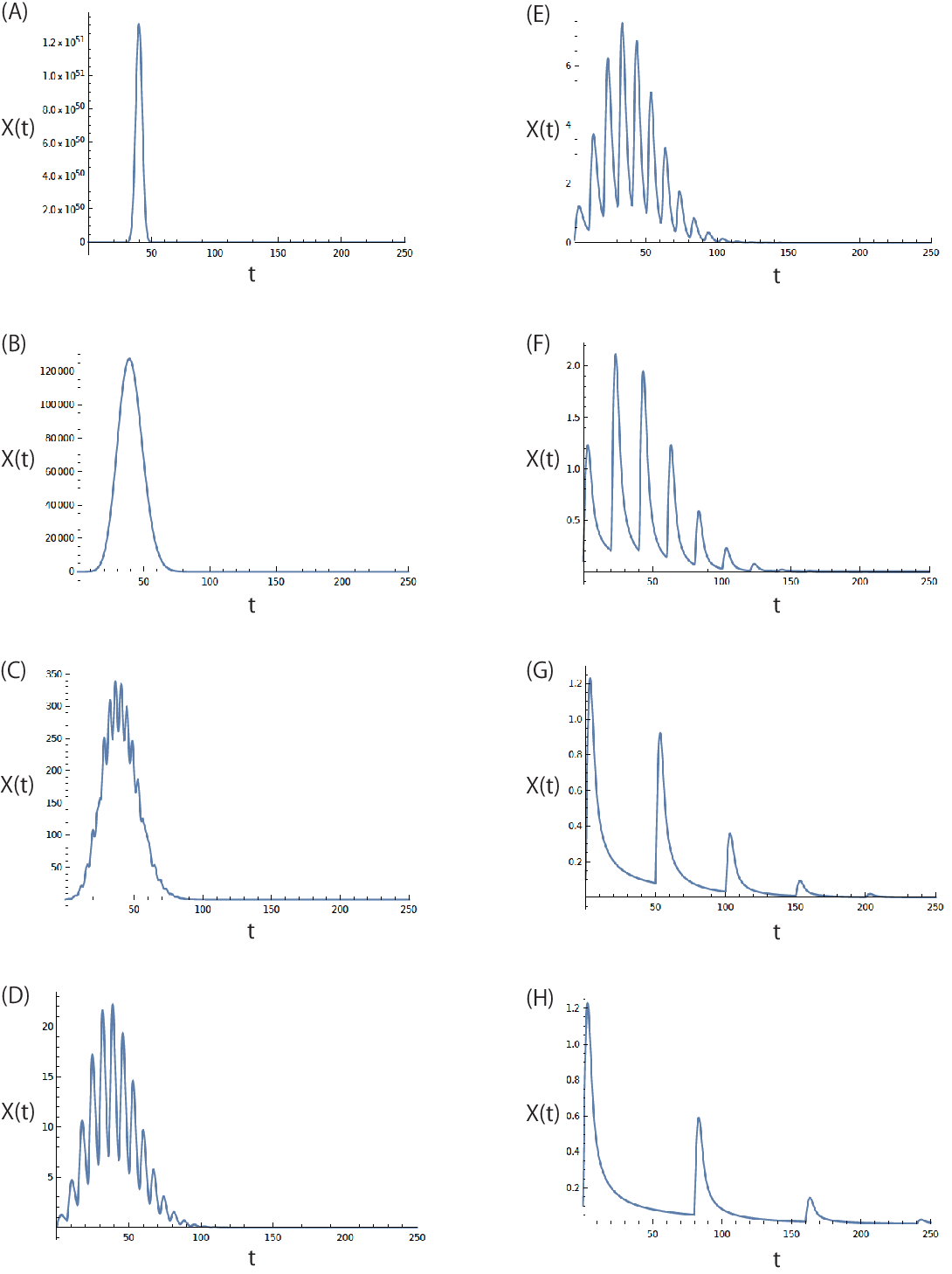}
\caption{Representative dynamics of the main equation (1) with different values of the delays, $\tau$. The parameters are set at $a=0.15, b= 6.0$ with the initial interval condition as
$X(t) = 0.1 (-\tau \leq t \leq 0)$. The values of the delays $\tau$ are (A)$0$, (B)$2$, (C)$4$, (D)$7$, (E)$10$, (F)$20$, (G)$50$, (H)$80$.}
\label{dynamics}
\end{figure}

Let us now analyze the oscillatory behavior. Using equation(\ref{ft2}), we can compute the power spectrum
as
\begin{equation}
S(\omega) = |\hat{X}(\omega)|^2 = \hat{X}(\omega)\hat{X}^*(\omega) = {\cal{C}}^2 Exp[ {- {1\over a} \omega^2 + {2 b\over \tau a} \cos{\omega \tau}}] 
\label{powerspectrum}
\end{equation}

\begin{figure}
\includegraphics[height=12cm]{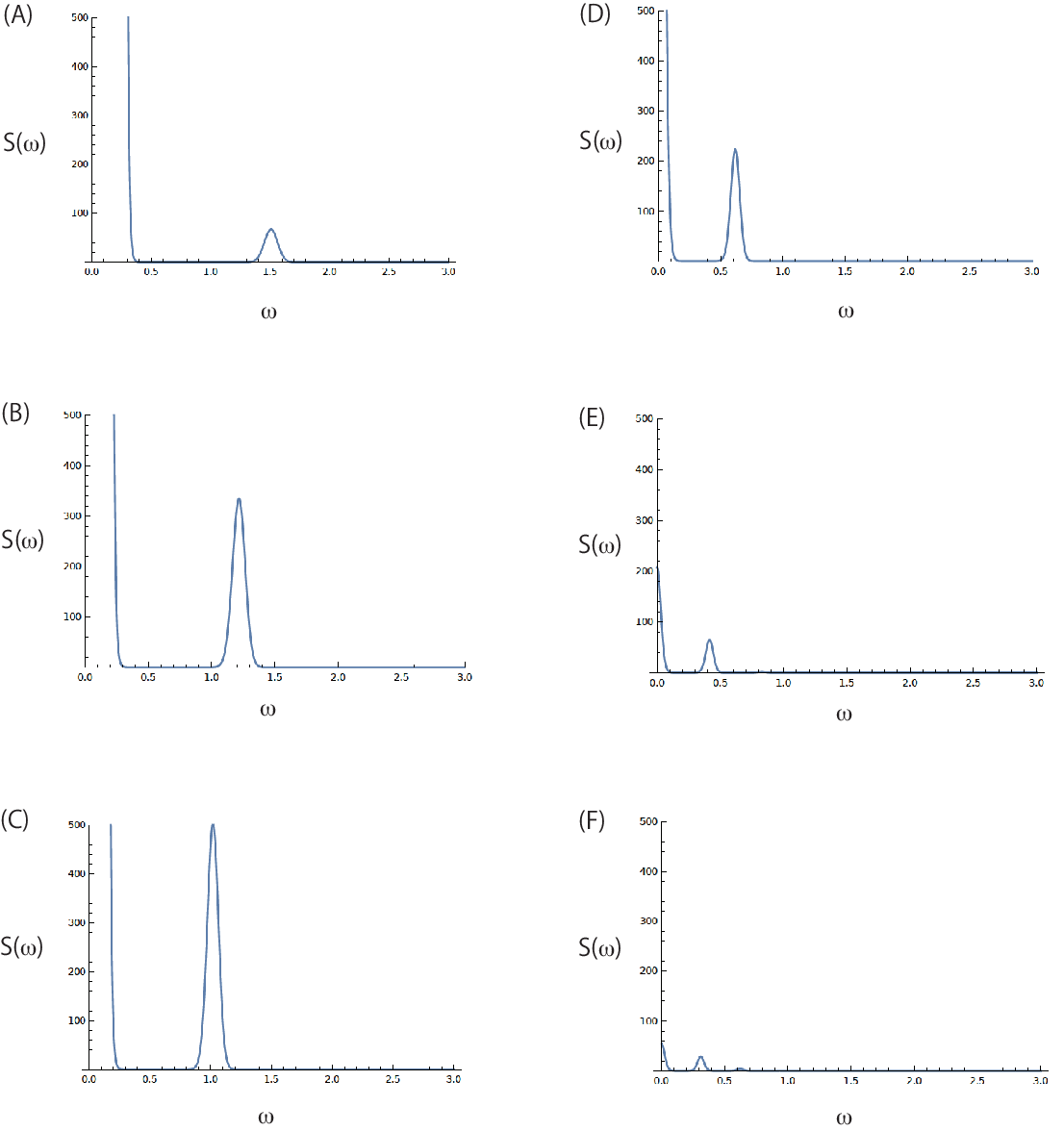}
\caption{Representative power spectrums given by the equation (1) with different values of the delays $\tau$. The parameters are set as the same as Fig.1; $a=0.15, b= 6.0, {\cal{C}}=1$ with the initial interval condition as
$X(t) = 0.1 (-\tau \leq t \leq 0)$. The values of the delays $\tau$ are (A)$4$, (B)$5$, (C)$6$, (D)$10$, (E)$15$, (F)$20$.}
\label{power}
\end{figure}
We have plotted this equation for the power spectrum for the various delays. Results with the same parameter setting as in Fig. 1 are shown in Fig. 2.

Also as shown, the peak of the power spectrum shows a maximum height with the tuned value of the delay. This indicates the resonance with the delay as a tuning parameter. We can analyze this by examining equation (\ref{powerspectrum}). By taking the derivative of (\ref{powerspectrum}), we see the maximum and minimum points of the power spectrum occurs at $\omega^*$ satisfying,
\begin{equation}
\omega^* = - b \sin{\omega^* \tau}
\end{equation}

The appearance and disappearance of oscillatory behavior correspond to that of the peaks in
the power spectrum.

They are given by the intersection points of the two functions from both sides of this condition (Fig.3(A-i); The value of $b$ is chosen to be different from Figs.1 and 2
so that the comparison is easier to be shown graphically.). The position of the first peak corresponds to the second non-zero smallest intersection point (the first one corresponds to the minimum before the peak). We can numerically estimate this point, and obtain the height
of the peak for various values of the delay, which is plotted in Fig.4. The resonance with the delay as the tuning parameter is clearly observed.

We can also infer the condition for the appearance of oscillatory behavior. From above, the critical value $\tau_c$ necessary for the existence of the power spectrum peaks are when the intersection point is the tangent point (Fig.3(B)). This gives the following conditions.
\begin{equation}
\omega^* = - b \sin{\omega^* \tau_c},\quad 1 = - b \tau_c \cos{\omega^* \tau_c}, \quad {\pi \over \tau_c} < \omega^* < {3\pi \over 2\tau_c}
\label{cond}
\end{equation}
If we set $\lambda_c = b \tau_c$, the above condition leads to
\begin{equation}
- {1 \over \lambda_c} = \cos{\sqrt{\lambda_c^2 -1}},\quad {\sqrt{\pi^2 + 1}} < \lambda_c < {\sqrt{{9\over 4}\pi^2 + 1}}
\end{equation}
We can obtain the value of $\lambda_c$ numerically as
\begin{equation}
 \lambda_c  = b \tau_c \approx 4.603.
\end{equation}
Thus, if we fix the parameter $b$, the oscillation appears for the delay longer than this critical delay:
\begin{equation}
\tau_c = {\lambda_c \over b}  < \tau.
\end{equation}
Also, if we increase the delay further, there are more peaks to appear as there are more roots to satisfy (\ref{cond}). Asymptotically, components of all frequency signals are mixed, leading to suppression of the oscillatory behavior.

\begin{figure}
\begin{center}
\includegraphics[height=9cm]{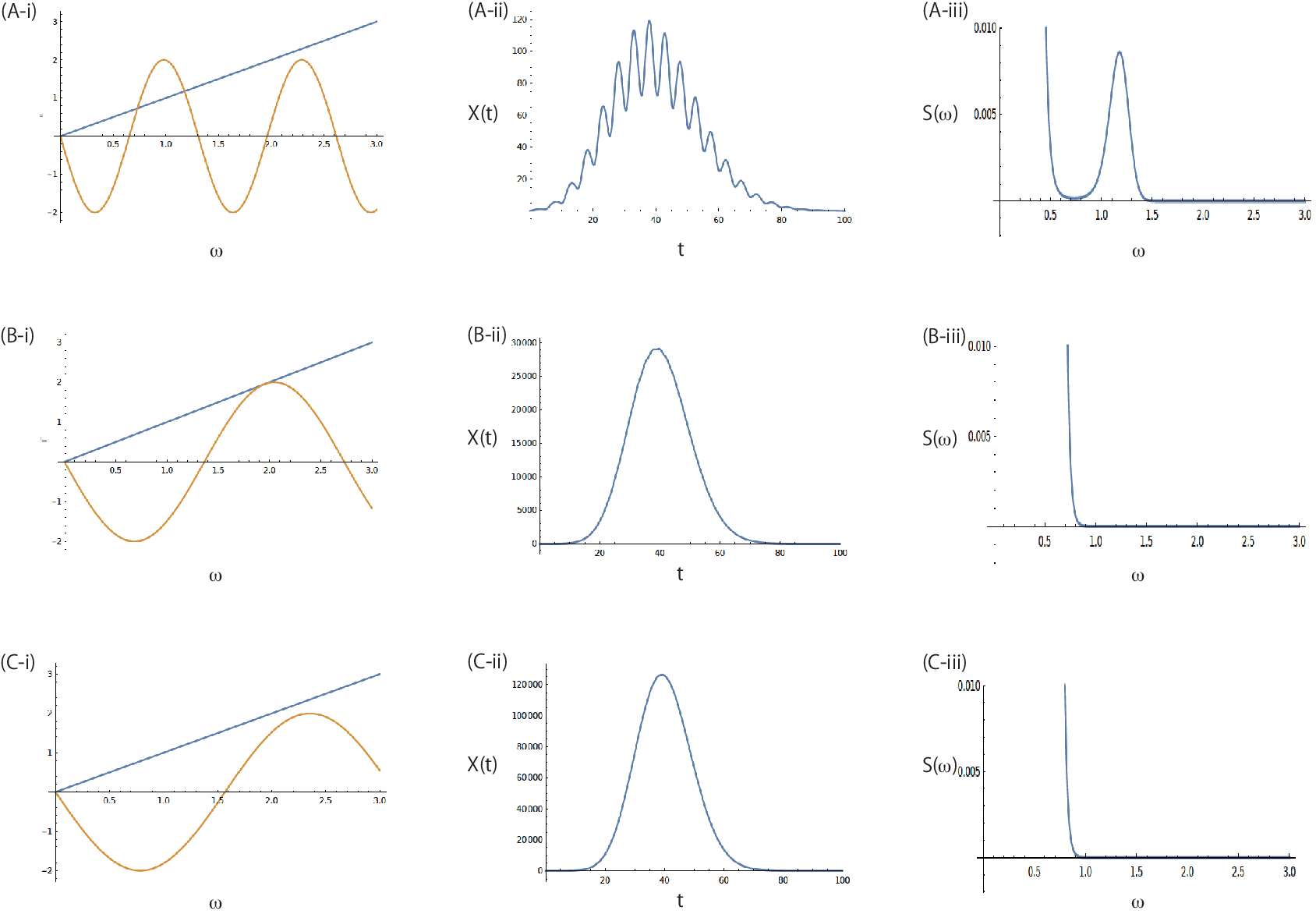}
\caption{Analysis of the peak position of the power spectrum (i), and the associated dynamics (ii) and the power spectrums (iii). The parameters are $a=0.15, b= 2.0$ with the initial interval condition as
$X(t) = 0.1 (-\tau \leq t \leq 0)$. The critical value of the delay is $\tau_c = {\lambda_c\over b} \approx 2.3017$ The values of the delays $\tau$ are (A)$4.802$, (B)$2.3017$, (C)$2.002$.}
\label{power}
\end{center}
\end{figure}

\begin{figure}
\begin{center}
\includegraphics[height=5cm]{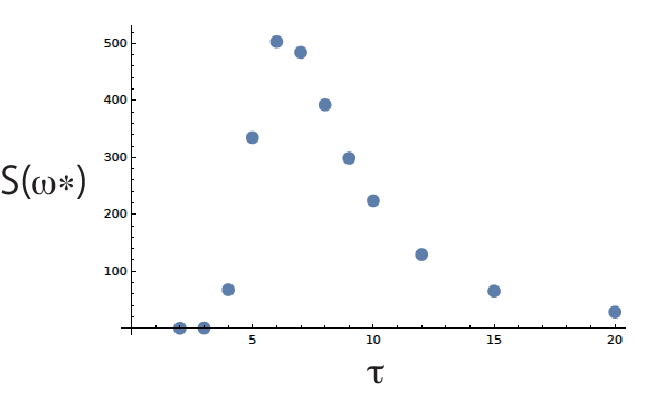}
\caption{Resonant curve with values of the delays $\tau$ as the tuning parameter. The parameters are set as the same as Fig.1; $a=0.15, b= 6.0$ with the initial interval condition as $X(t) = 0.1 (-\tau \leq t \leq 0)$.}
\label{resonant point}
\end{center}
\end{figure}

\clearpage

\section{Discussion}

In this paper, we proposed a simple delay differential equation that exhibits oscillatory resonance. Resonance with both stochasticity and delay have previously been investigated as 
``Delayed stochastic resonance''\cite{ohirasato1999}. There, the tuned combination of strength of 
noise and the amount of delay led to the more regular oscillatory patterns.  
Our equation here is simpler as it induces resonance only with the delay.
We can consider equations of the more general form,
\begin{equation}
{dX(t)\over dt} + a t X(t) = f( X(t-\tau) ),
\label{gdr}
\end{equation}
where the function $f$ can be a non-linear function. Our preliminary simulation results show that if we set $f$ in the form of the ``negative feedback function'' and the ``mixed feedback function''\cite{mackeyglass1977,glass1988,glassmackey1988}, we can observe transient oscillations and chaotic dynamics respectively. They are also suppressed with asymptotically larger delay. More investigations, however, are needed to understand the nature of these behaviors with the general equation.

\end{document}